\documentclass[twocolumn,preprintnumbers,superscriptaddress,endnote,nofootinbib,prl,letterpaper]{revtex4}
\usepackage{graphicx}
\usepackage{xcolor}

\newcommand{\vev}[1]{\langle {#1} \rangle}
\newcommand{\lsim}{\lesssim}
\newcommand{\gsim}{\gtrsim}

\newcommand{\eq}[1]{Eq.~(\ref{#1})}

\newcommand{\ord}[1]{\mathcal{O}{(#1)}}
\newcommand{\beq}{\begin{equation}}
\newcommand{\eeq}{\end{equation}}

\begin{document}

\pagestyle{plain}

\title{\boldmath A Tale of Two Anomalies}

\author{Hooman Davoudiasl
\footnote{email: hooman@bnl.gov}
}
\author{William J. Marciano
\footnote{email: marciano@bnl.gov}
}

\affiliation{Department of Physics, Brookhaven National Laboratory,
Upton, NY 11973, USA}


\begin{abstract}

A recent improved determination of the fine structure constant, $\alpha=
1/137.035999046(27)$, leads to a $\sim 2.4 \sigma$ negative discrepancy between the
measured electron anomalous magnetic moment and the Standard Model
prediction. That situation is to be compared with the muon anomalous
magnetic moment where a positive $\sim 3.7 \sigma$ discrepancy has existed for some
time.  A single scalar solution to both anomalies
is shown to be possible if the two-loop electron Barr-Zee diagrams dominate
the scalar one-loop electron anomaly effect and the scalar couplings to the electron and two photons are relatively large.  We also briefly discuss the implications of that scenario.

\end{abstract}
\maketitle


So far, neither the LHC experiments nor direct searches for dark matter (DM) have uncovered any
signs of a ``natural'' Higgs sector nor weak scale dark matter states.  However, there have been mild deviations from the Standard Model (SM) predictions over the years.  Of these, a long-standing one is the $\sim 3.7\, \sigma$ discrepancy between experiment \cite{Bennett:2006fi,PDG} and theory (see, for example, Refs.~\cite{Blum:2018mom,Keshavarzi:2018mgv}) for  the muon anomalous magnetic moment $a_\mu\equiv (g_\mu-2)/2$ 
\beq
\Delta a_\mu \equiv a_\mu^{\rm exp} - a_\mu^{\rm th} =  (274 \pm 73)\times 10^{-11},
\label{Delamu}
\eeq
which has withstood various theoretical refinements and is 
being currently remeasured at Fermilab with higher precision.  While the final word on $g_\mu-2$ remains 
to be decided by the new measurements and ongoing theoretical improvements of the SM prediction,  
the deviation has been a subject of intense phenomenological interest.     
As new physics at the TeV scale gets more constrained, the parameter space for weak scale 
models that could explain $g_\mu-2$ starts to close.  

Meanwhile, the search for new ``dark'' or ``hidden'' states at low mass scales $\lsim 1$~GeV 
has recently been getting increasing attention \cite{Bjorken:2009mm,Essig:2013lka}, partially spurred by astrophysical 
considerations related to DM models \cite{ArkaniHamed:2008qn} and perhaps also by the dearth of indications for new high energy phenomena.  In fact, $g_\mu -2$ has emerged as an interesting target for dark sector 
searches, since light states with feeble couplings to the SM can in principle explain the anomaly.  
An early and motivated possibility was offered by the ``dark photon'' hypothesis, where a new vector boson that kinetically mixes with the photon \cite{Holdom:1985ag} could have provided a 
solution \cite{Pospelov:2008zw}.  This idea and its simple extensions have now been essentially  ruled out.  However, other light states from a dark sector, 
for example a light scalar that very weakly couples to muons, could still furnish a potential solution 
\cite{Chen:2015vqy}.

A recent precise determination of the fine structure constant, $\alpha$, has introduced a new twist to this story.   An improvement in the measured $h/M_{\rm Cs}$ of
atomic Cesium, where $h$ is Planck's constant, used in conjunction with other precisely known mass ratios and the
Rydberg constant leads to the new best value \cite{Parkeretal} 
\beq
\alpha^{-1} (\text{Cs})= 137.035999046(27).
\label{Cs-alpha}
\eeq
(For a detailed explanation of that prescription and its use in determining
the SM prediction for the electron anomalous magnetic moment,
$a_e=(g_e-2)/2$, see the articles by G. Gabrielse in Ref.~\cite{Gabrielse}.)
As a result, 
comparison of the theoretical prediction of $a_e^{\rm SM}$ \cite{Aoyama:2017uqe} with the existing experimental measurement of $a_e^{\rm exp}$ \cite{Hanneke:2008tm,Hanneke:2010au} 
now leads to a discrepancy 
\begin{eqnarray}\label{DelaeDef}
\Delta a_e &\equiv& a_e^{\rm exp} - a_e^{\rm SM} \\
&=&  [-87 \pm28\, (\text{exp}) \pm 23 \, (\alpha) \nonumber 
\pm 2 \, (\text{\rm theory})]\\ \nonumber 
&\times& 10^{-14},
\end{eqnarray}
or when the uncertainties are added in quadrature
\beq
\Delta a_e = (-87 \pm 36)\times 10^{-14}.                                          
\label{Delae}
\eeq
The above result represents a $2.4\, \sigma$ discrepancy that is opposite in sign from the long standing muon
discrepancy previously mentioned and larger in magnitude than lepton-mass-scaling 
$m_e^2/m_\mu^2$ might suggest.  

Note that the discrepancy in Eqs.~(\ref{DelaeDef}) and 
(\ref{Delae}) results from an improvement
in $\alpha^{-1}$ from 137.035998995(85) which previously \cite{Aoyama:2017uqe} gave
$\Delta a_e =-130(77) \times 10^{-14}$ and represented a $1.7 \, \sigma$ effect. The central
value has decreased in magnitude, but its significance has increased.  The errors from the experimental determinations of $a_e$ and $\alpha$ are now the dominant sources of uncertainty 
and they are expected to further improve in the near future.  An alternative perspective is that $\alpha^{-1} (a_e)= 137.035999149(33)$, derived from a comparison of $a_e$ theory \cite{Aoyama:2017uqe} and experiment \cite{Hanneke:2008tm,Hanneke:2010au}, differs by $2.4 \, \sigma$ from \eq{Cs-alpha}, and follow-up experimental improvements may resolve the current discrepancy or significantly diminish its magnitude.  

Interestingly, dark photon models \cite{Pospelov:2008zw} and their simple extensions predict a one-loop positive deviation for both $g_\mu-2$ and $g_e-2$.  Therefore, the negative $\sim 2.4\,\sigma$ deviation in $g_e-2$ cannot be simultaneously 
explained together with the $\sim 3.7\,\sigma$ anomaly in $g_\mu-2$ in the simplest versions of those models, even if one could circumvent existing experimental constraints.   

In this paper, we would like to point out that a minimal model based on a single light 
real scalar $\phi$, can in principle explain the deviations of both $g_\mu-2$ and $g_e-2$, in a relatively economical fashion.  We will show that a two-loop Barr-Zee diagram \cite{Barr:1990vd,Bjorken:1977vt} might explain $\Delta a_e$ 
while a one-loop contribution could be the primary origin of $\Delta a_\mu$ \cite{Chen:2015vqy,Batell:2016ove}, with both corrections mediated by the same scalar $\phi$.  
For more detailed discussions of these loop processes and their contributions  
to the electron and muon anomalous magnetic moments see 
Ref.~\cite{Giudice:2012ms,Marciano:2016yhf}, where the authors discuss the relative contributions of one- and
two-loop diagrams, but focus primarily on the case of a pseudoscalar boson.  

Before going further, we note that somewhat less minimal solutions, {\it e.g.} with a scalar coupled to the muon and a pseudo-scalar coupled to the electron, can potentially yield the right size and sign for the deviations in $g_\mu-2$ and $g_e-2$, respectively, and satisfy experimental constraints. However, here, we focus on the effect of a single light scalar where inclusion of 
the Barr-Zee contribution represents an extension of earlier work in 
Ref.~\cite{Chen:2015vqy}.  Studies of the contribution of Barr-Zee type diagrams 
to $g_\mu-2$ in the context of two Higgs doublet models and supersymmetry can also be found in 
Ref.~\cite{Cheung:2001hz}.  

Let us consider the following effective Lagrangian for the real scalar $\phi$ of mass $m_\phi$
\beq
{\cal L_\phi} = -\frac{1}{2} m_\phi^2 \phi^2 - \sum_f\, \lambda_f \phi \,\bar f f 
- \frac{\kappa_\gamma}{4} \, \phi \, F_{\mu \nu}F^{\mu\nu},
\label{Lphi}
\eeq 
where we only include explicit couplings with strengths $\lambda_f$ to a set of fermions $f$ and have omitted various kinetic terms and fermion masses.  In this work, we allow $f$ to correspond to SM fermions, 
as well as other potential more massive charged fermions.  The $\lambda_f$ are constrained by 
phenomenology, as will be discussed later.  We 
assume that the $\phi$ coupling to photons, through the field strength tensor $F_{\mu\nu}$, is governed by the the constant $\kappa_\gamma$ which has mass dimension $-1$.  
The sum over $\phi \gamma \gamma$ triangle diagrams mediated by $f$ will induce 
a contribution to $\kappa_\gamma$, 
but we do not specify the properties of all charged states that couple to $\phi$. 

\begin{figure}
\includegraphics[width=0.42\textwidth]{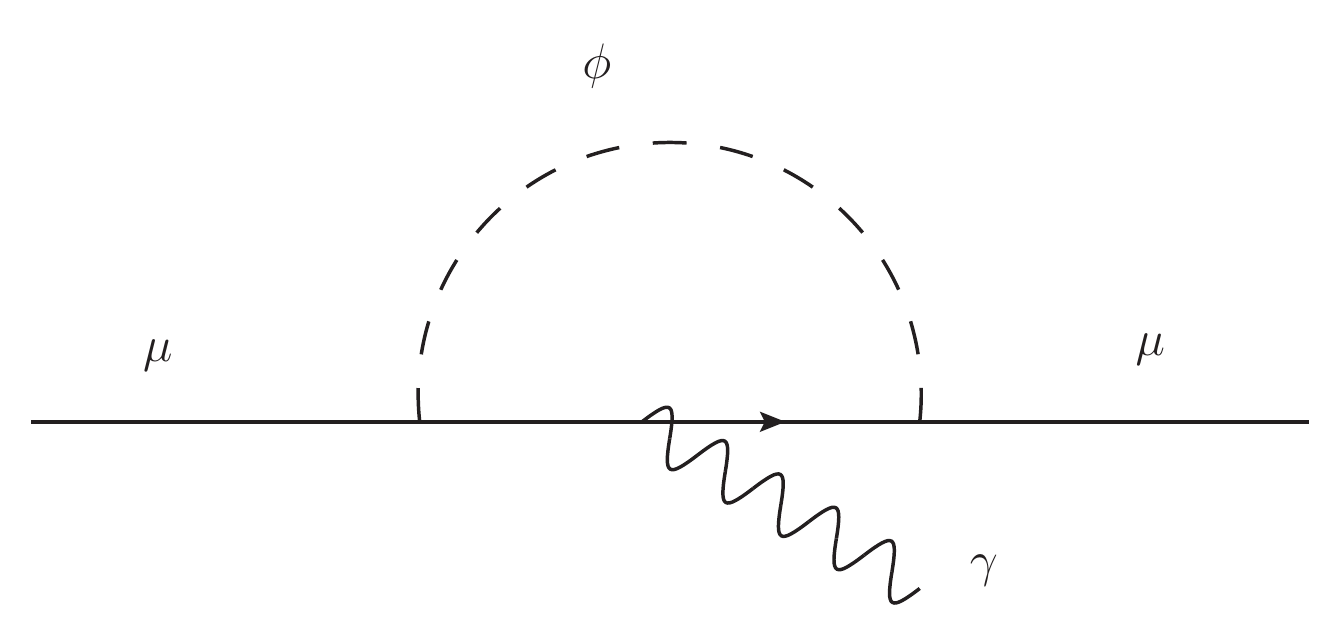}
 \caption{One-loop $\phi$ contribution to $g_\mu-2$.}
 \label{1loop}
\end{figure}

We will start with the $g_\mu-2$ discrepancy, assumed to be 
dominated by the one-loop diagram in Fig.\ref{1loop},  
which is given by \cite{Chen:2015vqy,Leveille:1977rc,TuckerSmith:2010ra} 
\beq
\Delta a_\ell = \frac{\lambda_\ell^2}{8 \pi^2}\, x^2 \int_0^1 d z \frac{(1+z)(1-z)^2}{x^2(1-z)^2 + z}\,
\label{Delaell}
\eeq
for a lepton $\ell$ of mass $m_\ell$ and $x\equiv m_\ell/m_\phi$.

Current experimental constraints, as illustrated in Ref.~\cite{Batell:2017kty} - under the assumption that $\phi$ only couples to muons - allow $2 m_\mu \lsim m_\phi \lsim 100$~GeV and $\lambda_\mu \sim 5 \times 10^{-4} - 0.1$, roughly corresponding to a range of parameters that can explain the 3.7$\sigma$ deviation in $g_\mu-2$, given by \eq{Delamu}, which we will 
approximate as $\Delta a_\mu \approx 3\times 10^{-9}$.  The above lower bound on $m_\phi$ corresponds to demanding that $\phi$ decay promptly into muon pairs.  In our scenario, couplings to the electron lead to prompt decays $\phi \to e^+ e^-$ below the muon pair threshold, allowing $m_\phi \lsim 200$~MeV.  However, for such values of $m_\phi$, the one-loop positive contribution to $g_e-2$ starts to become significant and cancel out the desired two-loop effect that we will discuss below.  For $m_\phi$ well above the GeV scale, we also find it difficult to accommodate the suggested $g_e-2$ anomaly in \eq{Delae} with reasonable values of $\lambda_e$ and $\kappa_\gamma$.  In addition, for $m_\phi\gg 1$~GeV, typical low energy probes of $\phi$ at intense beam facilities become less efficient, adversely affecting experimental prospects for testing the scenario.  For the above reasons, we mostly focus on the $\phi$ mass range $2 m_\mu \lsim m_\phi \lsim \text{few}$~GeV, in what follows.

Let us choose, for concreteness, 
\beq
m_\phi = 250~\text{MeV} \quad \text{and} \quad \lambda_\mu = 10^{-3}, 
\label{refpar}
\eeq
which according to \eq{Delaell} yields 
$\Delta a_\mu \approx 3\times 10^{-9}$.  

\begin{figure}
\includegraphics[width=0.42\textwidth]{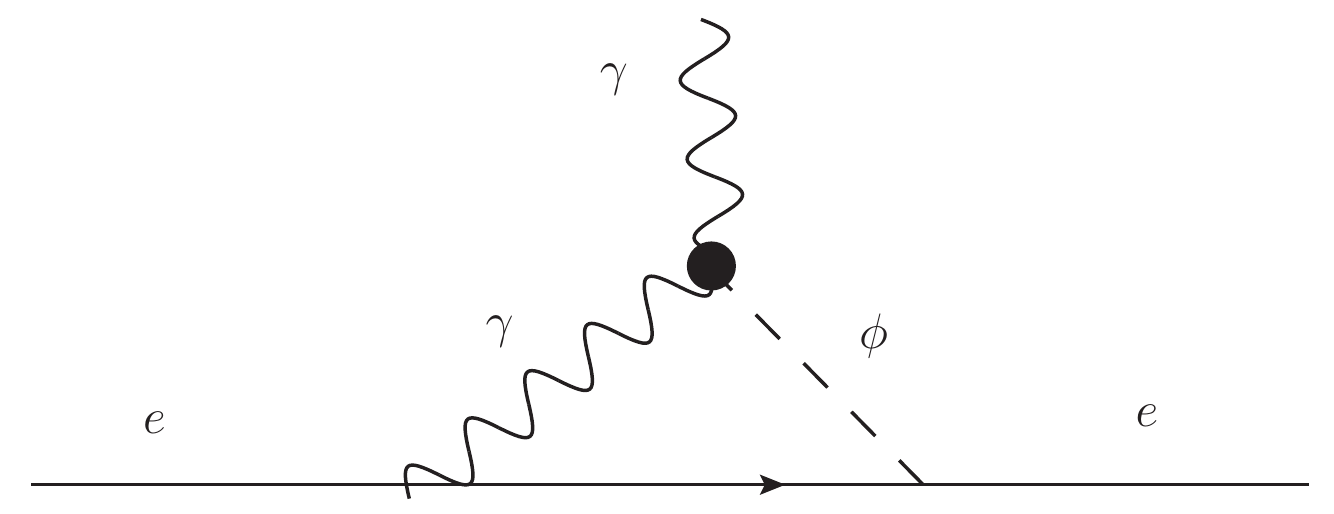}
 \caption{Effective two-loop Barr-Zee diagram contribution to $g_e-2$, with fermion loops integrated out.  The dot ($\bullet$) represents light and heavy fermion 
 loops that contribute to $\kappa_\gamma$.}
 \label{BZ}
\end{figure}
We now address the deviation in \eq{Delae}.  Here, we will concentrate on the ``Barr-Zee'' 
diagram contribution to $a_\ell$ in Fig.\ref{BZ}, for a heavy fermion $f$ loop that is represented by the dot ($\bullet$) in the figure, 
given by \cite{Barr:1990vd,Giudice:2012ms}
\beq
\Delta a_\ell^{\rm BZ}(f) = -\frac{\alpha}{6 \pi} \frac{m_\ell}{m_f} \frac{\lambda_\ell \lambda_f}{\pi^2} Q_f^2 N_c^f\, I(y)\,,
\label{2loop}
\eeq
where 
\beq
I(y) = \frac{3}{4} y^2 \int_0^1dz \frac{1 - 2 z(1-z)}{z(1-z) - y^2}\ln \frac{z(1-z)}{y^2}, 
\label{Iy}
\eeq
with $y\equiv m_f/m_\phi$; $Q_f$ and $N_c^f$ are the electric charge and the number of colors of $f$, respectively, with $N_c^f = 1(3)$ for ordinary leptons (quarks).  
For multiple heavy fermions, one simply sums over $f$.

For $y^2\gg 1$, the above expression for $\Delta a_\ell^{\rm BZ}$ will reduce to \cite{Giudice:2012ms,Czarnecki:2017rlm}  
\beq
\Delta a_\ell^{\rm BZ} \approx 
\frac{\lambda_\ell\, \kappa_\gamma\, m_\ell}{4\,\pi^2} (13/12 + \ln y),
\label{DelaellBZ}
\eeq
after integrating out heavy charged fermions of mass $m_f$ in the two-loop Barr-Zee diagram.  Here, $\kappa_\gamma$ is given by  (see, for example, Ref.~\cite{Carena:2012xa}) 
\beq
\kappa_\gamma \approx - \frac{2\, \alpha}{3 \pi}\sum_f \frac{\lambda_f \, Q_f^2 \, N^f_c}{m_f},  
\label{kappa}
\eeq  
where it is assumed that the sum is over fermions with similar values of
$\ln(m_f/m_\phi)$.  Otherwise, the terms in \eq{kappa} should be weighted by the different values of $I(m_f/m_\phi)$.  The above formula for $\kappa_\gamma$ is obtained 
in the limit that $y^2\gg 1$.  For heavy fermions to contribute significantly to $\kappa_\gamma$ they need to couple to $\phi$ with sizable strength.

It is instructive to consider various potential contributions to $\Delta a_e^{\rm BZ}$  provided by $\kappa_\gamma$ 
values and their associated $I(m_f/m_\phi)$ for $m_\phi=250$~MeV.  
We consider the relative muon, tau and generic TeV particle BZ contributions, assuming $\lambda_e=4\times 10^{-4}$, roughly the maximum value allowed by ``dark photon'' 
searches \cite{Lees:2014xha}.  In terms of their possible $\lambda_f$ values, one finds
the following relative BZ contributions in units of $10^{-14}$: $-5\, \lambda_\mu/10^{-3}, 
-14\, \lambda_\tau/10^{-2}$, and $-7.5\, \lambda_{\rm TeV}$.
The muon value $\lambda_\mu= 10^{-3}$ is fixed by $\Delta a_\mu$ and is only capable of accounting for a small, $\sim6\%$ ,of the central discrepancy in \eq{Delae}. The tau
has more $\lambda_\tau$ freedom. A rather large, but not ruled out, value
of $\lambda_\tau=0.06$ could accommodate the entire $\Delta a_e$ discrepancy.  
Of course, such a large $\lambda_\tau$ would have many other phenomenological consequences for tau physics.  In the case of new TeV
charged states, about 10 new states with $\ord{1}$ Yukawa couplings are 
required to account for the discrepancy.  
The TeV scenario would seem most viable in a theory with dynamical symmetry breaking.  Some combination of the above contributions  
with smaller couplings could also achieve the desired value of $\Delta a_e$.  
For larger $m_\phi$ values, in the $\ord{\rm GeV}$ regime, similar conclusions can be reached.

Let us focus on the contributions of the tau\footnote{We will not consider couplings of $\phi$ to quarks, in order to avoid potentially severe constraints from flavor-changing-neutral currents.  Given the current 
mild $B$ physics anomalies pointing to lepton flavor universality violations such an extension may be worth further examination.  However, that analysis is outside the scope of this paper whose focus is on the leptonic interactions of the scalar $\phi$.}
or TeV scale fermions, and for concreteness take 
\beq
\lambda_e = 3\times 10^{-4} \,,
\label{refpar2}
\eeq
which is somewhat below the aforementioned experimental upper bound.  To account for the $g_e-2$ discrepancy in \eq{Delae}, we then require $\phi$-photon coupling to be 
\beq
\kappa_\gamma = - 7 (2) \times10^{-5}~ 
\text{GeV}^{-1} \; \text{for}\; m_f = m_\tau \; (\sim 1~\text{TeV}).
\label{kappaval}
\eeq
Here, we have assumed that the sign of the $\kappa_\gamma$ is negative, which is a 
choice corresponding to positive fermion Yukawa couplings to $\phi$.   
Assuming a coupling strength $\lambda_\tau \sim 8\times 10^{-2}$ of $\phi$ 
to $\tau$, \eq{kappa} would yield $\kappa_\gamma \sim - 7\times 10^{-5}$.   We note that a more careful phenomenological study is perhaps required to determine whether a coupling of $\ord{8\times 10^{-2}}$ between $\phi$ and $\tau$ is consistent with experimental constraints. 
Thus, we generally expect that contributions from electrically charged states of mass $\gsim \text{few}\times 100$~GeV (the new charged 
states cannot be  much lighter given the fair agreement of the TeV scale LHC data with the SM predictions) are needed to generate the 
requisite strength of $\kappa_\gamma$.   As noted before, to account for the entire $g_e-2$  discrepancy 
would require constructive contribution of $\ord{10}$ 
fermions at the TeV scale with $\lambda_f \sim 1$.  This requirement can be moderated if the tau contribution is significant, say with $\lambda_\tau \sim \text{few}\times 10^{-2}$, assuming constructive interference from same sign Yukawa couplings.  

Note that the chosen value of $\lambda_e$ in \eq{refpar2} does not follow naive scaling with the lepton mass, {\it i.e.} $\lambda_e/\lambda_\mu = m_e/m_\mu$, in reference to that of $\lambda_\mu$ in \eq{refpar}.  However, since $\phi$ is not assumed to control the masses of the leptons, this is not an inconsistent choice and can be easily obtained from 
a simple effective theory that does not have hierarchic charged lepton interactions.  We would also like to mention that for values of $\ln(m_f/m_\phi)$ larger than those assumed in the preceding discussion, one could choose smaller values of $|\lambda_{e,\mu}|$ due to the enhanced contributions of the Barr-Zee diagrams to both $a_e$ and $a_\mu$, for 
$\lambda_e \,\kappa_\gamma <0$ and $\lambda_\mu \,\kappa_\gamma >0$, respectively.  This would presumably originate  from the coupling of $\phi$ to the aforementioned charged heavy states.  We will see later that such new particles may be motivated in some ultraviolet completions of 
the effective theory in \eq{Lphi}.

For the above reference values, the one-loop contribution 
from \eq{Delaell} to $\Delta a_e$ is $\sim 5\times 10^{-14}$, 
which is small compared to that required by the apparent anomaly in \eq{Delae}.  Similarly, 
the Barr-Zee diagram contribution to $a_\mu$ from \eq{DelaellBZ} is $\sim - 5 \times 10^{-10}$, 
a factor of $\sim 5$ too small and of the wrong sign to account for the anomaly in $a_\mu$ from \eq{Delamu}. To compensate for this $\sim 20\%$ effect we could change our reference parameters in \eq{refpar} slightly, but the values chosen here suffice to illustrate that, in principle, a simultaneous resolution of both current discrepancies in 
$a_e$ and $a_\mu$ can be obtained in our framework.
We also note that at larger $m_\phi$ alternative possibilities arise.  For example, 
for $m_\phi = 3.0$~GeV, using a value of $\kappa_\gamma$ that gives a Barr-Zee two-loop solution 
for $\Delta a_e$ with $\lambda_e = 3.0 \times 10^{-4}$, we find that there are two values of $\lambda_\mu \sim -3 \times 10^{-3}$ and $\sim 1\times 10^{-2}$ that yield the $g_\mu-2$ discrepancy, from the sum of one- and two-loop diagrams.  Here, the two-loop contribution for the negative $\lambda_\mu$ is dominant over that of the one-loop diagram, and vice versa for the positive $\lambda_\mu$.  

Aspects of phenomenology related to the coupling of $\phi$ to muons, including extensions 
to CP violating couplings, have been discussed before \cite{Chen:2015vqy,Batell:2016ove}.  The coupling of light $\phi$ to leptons could lead to signals for ``bump hunt'' searches in decay or scattering processes, if kinematically allowed.  
Assuming significant $\phi$ coupling to $\tau$ of $\lambda_\tau \gsim \text{few}\times 10^{-2}$, 
 leptonic ($l=e,\mu$) resonances in $\tau$ decay, for example $\tau \to e \,\nu \,\bar \nu \,\phi (\to l^+l^-)$, 
 or in scattering $e^+ e^- \to \tau^+ \tau^- \phi (\to l^+ l^-)$, as suggested in 
 Ref.~\cite{Batell:2016ove}, could provide potentially promising signals.

Regardless of the production process for $\phi$, its dominant decay modes play an important role in 
its phenomenology and experimental implications.  Let us first consider the decay of $\phi$ into an on-shell  fermion pair $\bar f f$ that interact with it through the Yukawa coupling in \eq{Lphi}.  The partial width for this decay is given by 
\beq
\Gamma(\phi\to \bar f f) = \frac{\lambda_f^2}{8 \pi}m_\phi\left(1 - \frac{4 m_f^2}{m_\phi^2}
\right)^{3/2}.
\label{Gamff}
\eeq
The decay of $\phi$ into photons, assuming the coupling $\kappa_\gamma$ in \eq{Lphi}, is 
given by 
\beq
\Gamma(\phi\to \gamma \gamma) = \frac{\kappa_\gamma^2}{64 \pi}m_\phi^3.
\label{Gamff}
\eeq
(See, for example, Ref.~\cite{Batell:2017kty}.)  
The coupling $\kappa_\gamma$ is generated by the interactions of $\phi$ with charged fermions as previously discussed.  Here, rather than specify all such charged states, we parametrize the $\phi \gamma \gamma$ overall interaction in terms of the effective coupling $\kappa_\gamma$.  We find that for the above chosen reference values (\ref{refpar}), (\ref{refpar2}), and (\ref{kappaval}) 
we get $\Gamma(\phi\to \mu^+ \mu^-) \approx 1.5$~eV, 
$\Gamma(\phi\to e^+ e^-) \approx 0.9$~eV,  $\Gamma(\phi\to \gamma \gamma) \approx (3-40) \times 10^{-5}$~eV, hence $\phi$ decays into muons and electrons with branching fractions of $\sim 60\%$ and $\sim 40\%$, respectively.  However, as $m_\phi$ gets larger than 250 MeV, the phase space suppression for the $\mu^+\mu^-$ final state becomes less important and the ratio of 
those branching fractions approaches $\sim 11$, for our chosen values of $\lambda_\mu$, 
$\lambda_e$, and $m_\phi \lsim 1$~GeV.

Let us briefly discuss potential models that could give rise to the types of interactions we have assumed.  The couplings of $\phi$ to $\mu$ and $e$ can be obtained from an 
effective operator of the form 
\beq
c_\ell\frac{\phi\, H \, \bar L\, \ell_R}{M}\,
\label{EO}
\eeq
where $M$ is the typical scale of new physics leading to the effective interaction above, 
$L$ is a lepton doublet of the SM, and $\ell_R$ is a right-handed charged lepton.  To avoid constraints from flavor changing neutral current data, we generally assume that the structure of these interactions are flavor-diagonal. Also, for typical 
parameters similar to our reference values assumed before, 
the underlying interactions generating the above operators are roughly flavor universal, that is 
$c_e \sim c_\mu$.  We then have $\lambda_\ell \equiv c_\ell \vev{H}/M$. Assuming 
$M\sim 1$~TeV, we find that $c_\ell \lsim 10^{-2}$.  

The scale $M$ in \eq{EO} could be identified with the mass of a vector-like fermion $F$, with quantum numbers of $\ell_R$.   
In Ref.~\cite{Chen:2015vqy} a similar setup was assumed, where lepton 
flavor violation constraints and model building issues were discussed in more detail.  Here, we simply take $\phi$ to be a singlet and not responsible for ``dark'' gauge symmetry breaking as was done in Ref.~\cite{Chen:2015vqy}.  Thus, couplings of the form $y_H H\bar L F_R$ and $y_\phi \phi \bar F_L \ell_R$ can generate the operator in \eq{EO}, 
with $c_\ell = y_H \, y_\phi$.  We note that this choice of ultraviolet theory is also consistent with 
the assumption of charged TeV scale particle contributions to the $\phi\gamma\gamma$ coupling $\kappa_\gamma$, discussed earlier.     
 
In summary, we have shown that a simple model, comprising a singlet scalar $\phi$ of mass $\gsim 250$~MeV and couplings $\sim 10^{-3}$ and $\text{few} \times 10^{-4}$ to the muon and the electron, 
respectively, can potentially account for a 
$\sim 3.7\, \sigma$ discrepancy in the muon $g-2$ and the $\sim 2.4\,\sigma$ discrepancy 
(or a significant part thereof),  
in the electron 
$g-2$ of opposite sign.  For $m_\phi\sim 250$~MeV, the former anomaly is mediated through a one-loop digram, whereas the latter originates mostly from a two-loop 
Barr-Zee diagram, using a phenomenologically allowed coupling of the 
scalar to photons as small as ${\rm few}\times 10^{-5}$~GeV$^{-1}$.  
Variations on this scenario, 
where two-loop contributions to the muon $g-2$ are important or dominant, can arise for $m_\phi \gsim 1$~GeV. 
The model could give rise to lepton pair signals in rare meson or tau decays, 
as well as those in electron and muon scattering processes.  A simple effective theory that does not lead to naive scaling of the scalar-lepton couplings with the lepton mass can realize our scenario.  The effective theory, in turn, could arise from TeV-scale charged vector-like fermions coupled to the SM Higgs and $\phi$.  In that case, the LHC could potentially discover those fermions, which would shed further light on the underlying physics manifested in the possible deviations of the electron and muon $g-2$.

\begin{acknowledgments}

Work supported by the US Department of Energy under Grant Contract DE-SC0012704.
\end{acknowledgments}



\begin{thebibliography}{99}

\bibitem{Bennett:2006fi}
  G.~W.~Bennett {\it et al.}  [Muon G-2 Collaboration],
  Phys.\ Rev.\ D {\bf 73}, 072003 (2006)
  [hep-ex/0602035].


\bibitem{PDG} 
  C.~Patrignani {\it et al.} [Particle Data Group],
  Chin.\ Phys.\ C {\bf 40}, no. 10, 100001 (2016).
  doi:10.1088/1674-1137/40/10/100001
  
\bibitem{Blum:2018mom}
  T.~Blum {\it et al.} [RBC and UKQCD Collaborations],
  arXiv:1801.07224 [hep-lat].
  
\bibitem{Keshavarzi:2018mgv} 
  A.~Keshavarzi, D.~Nomura and T.~Teubner,
  arXiv:1802.02995 [hep-ph].
  
  
\bibitem{Bjorken:2009mm}
  J.~D.~Bjorken, R.~Essig, P.~Schuster and N.~Toro,
  Phys.\ Rev.\ D {\bf 80}, 075018 (2009)
  [arXiv:0906.0580 [hep-ph]].
  
\bibitem{Essig:2013lka}
  R.~Essig, J.~A.~Jaros, W.~Wester, P.~H.~Adrian, S.~Andreas, T.~Averett, O.~Baker and B.~Batell {\it et al.},
  arXiv:1311.0029 [hep-ph].
  
\bibitem{ArkaniHamed:2008qn}
  N.~Arkani-Hamed, D.~P.~Finkbeiner, T.~R.~Slatyer and N.~Weiner,
  Phys.\ Rev.\ D {\bf 79}, 015014 (2009)
  [arXiv:0810.0713 [hep-ph]].
  
\bibitem{Holdom:1985ag} 
  B.~Holdom,
  Phys.\ Lett.\  {\bf 166B}, 196 (1986).
  doi:10.1016/0370-2693(86)91377-8
  
\bibitem{Pospelov:2008zw}
  M.~Pospelov,
  Phys.\ Rev.\ D {\bf 80}, 095002 (2009)
  [arXiv:0811.1030 [hep-ph]].
  
\bibitem{Chen:2015vqy} 
  C.~Y.~Chen, H.~Davoudiasl, W.~J.~Marciano and C.~Zhang,
  Phys.\ Rev.\ D {\bf 93}, no. 3, 035006 (2016)
  doi:10.1103/PhysRevD.93.035006
  [arXiv:1511.04715 [hep-ph]].

\bibitem{Parkeretal}
R.~H.~Parker, C. Yu, W. Zhong, B. Estey, H. Mueller, 
Science 360, 191 (2018).

\bibitem{Gabrielse}
G.~Gabrielse in Lepton Dipole Moments, edited by B.~Lee
Roberts and William~J.~Marciano, World Scientific.

\bibitem{Aoyama:2017uqe} 
  T.~Aoyama, T.~Kinoshita and M.~Nio,
  Phys.\ Rev.\ D {\bf 97}, no. 3, 036001 (2018)
  doi:10.1103/PhysRevD.97.036001
  [arXiv:1712.06060 [hep-ph]].
  
\bibitem{Hanneke:2008tm} 
  D.~Hanneke, S.~Fogwell and G.~Gabrielse,
  Phys.\ Rev.\ Lett.\  {\bf 100}, 120801 (2008)
  doi:10.1103/PhysRevLett.100.120801
  [arXiv:0801.1134 [physics.atom-ph]].
  
\bibitem{Hanneke:2010au} 
  D.~Hanneke, S.~F.~Hoogerheide and G.~Gabrielse,
  Phys.\ Rev.\ A {\bf 83}, 052122 (2011)
  doi:10.1103/PhysRevA.83.052122
  [arXiv:1009.4831 [physics.atom-ph]].
  
\bibitem{Barr:1990vd} 
  S.~M.~Barr and A.~Zee,
  Phys.\ Rev.\ Lett.\  {\bf 65}, 21 (1990)
  Erratum: [Phys.\ Rev.\ Lett.\  {\bf 65}, 2920 (1990)].
  doi:10.1103/PhysRevLett.65.2920, 10.1103/PhysRevLett.65.21
  
\bibitem{Bjorken:1977vt} 
  J.~D.~Bjorken and S.~Weinberg,
  Phys.\ Rev.\ Lett.\  {\bf 38}, 622 (1977).
  doi:10.1103/PhysRevLett.38.622
  
  
\bibitem{Batell:2016ove} 
  B.~Batell, N.~Lange, D.~McKeen, M.~Pospelov and A.~Ritz,
  Phys.\ Rev.\ D {\bf 95}, no. 7, 075003 (2017)
  doi:10.1103/PhysRevD.95.075003
  [arXiv:1606.04943 [hep-ph]].
  
  
\bibitem{Giudice:2012ms} 
  G.~F.~Giudice, P.~Paradisi and M.~Passera,
  JHEP {\bf 1211}, 113 (2012)
  doi:10.1007/JHEP11(2012)113
  [arXiv:1208.6583 [hep-ph]].
 
\bibitem{Marciano:2016yhf} 
  W.~J.~Marciano, A.~Masiero, P.~Paradisi and M.~Passera,
  Phys.\ Rev.\ D {\bf 94}, no. 11, 115033 (2016)
  doi:10.1103/PhysRevD.94.115033
  [arXiv:1607.01022 [hep-ph]].

  
\bibitem{Cheung:2001hz} 
  K.~m.~Cheung, C.~H.~Chou and O.~C.~W.~Kong,
  Phys.\ Rev.\ D {\bf 64}, 111301 (2001)
  doi:10.1103/PhysRevD.64.111301
  [hep-ph/0103183].
 
  
\bibitem{Leveille:1977rc} 
  J.~P.~Leveille,
  Nucl.\ Phys.\ B {\bf 137}, 63 (1978).
  doi:10.1016/0550-3213(78)90051-2
  
\bibitem{TuckerSmith:2010ra} 
  D.~Tucker-Smith and I.~Yavin,
  Phys.\ Rev.\ D {\bf 83}, 101702 (2011)
  doi:10.1103/PhysRevD.83.101702
  [arXiv:1011.4922 [hep-ph]].

\bibitem{Batell:2017kty} 
  B.~Batell, A.~Freitas, A.~Ismail and D.~Mckeen,
  arXiv:1712.10022 [hep-ph].
  

\bibitem{Czarnecki:2017rlm} 
  A.~Czarnecki and W.~J.~Marciano,
  Phys.\ Rev.\ D {\bf 96}, no. 11, 113001 (2017)
  Erratum: [Phys.\ Rev.\ D {\bf 97}, no. 1, 019901 (2018)]
  doi:10.1103/PhysRevD.97.019901, 10.1103/PhysRevD.96.113001
  [arXiv:1711.00550 [hep-ph]].
  
\bibitem{Carena:2012xa} 
  M.~Carena, I.~Low and C.~E.~M.~Wagner,
  JHEP {\bf 1208}, 060 (2012)
  doi:10.1007/JHEP08(2012)060
  [arXiv:1206.1082 [hep-ph]].
  
\bibitem{Lees:2014xha} 
  J.~P.~Lees {\it et al.} [BaBar Collaboration],
  Phys.\ Rev.\ Lett.\  {\bf 113}, no. 20, 201801 (2014)
  doi:10.1103/PhysRevLett.113.201801
  [arXiv:1406.2980 [hep-ex]].

\end{thebibliography}
\end{document}